\documentclass[manuscript,screen]{acmart}
\usepackage{amsmath}
\usepackage{graphicx}
\usepackage{subcaption}
\usepackage{url}
\usepackage[font=small,skip=3pt]{caption}

\AtBeginDocument{%
  \providecommand\BibTeX{{%
    \normalfont B\kern-0.5em{\scshape i\kern-0.25em b}\kern-0.8em\TeX}}}

\setcopyright{none}
\copyrightyear{2023}
\acmYear{2023}
\acmDOI{}

\acmConference[FAccTRec '23]{6th FAccTRec Workshop: Responsible Recommendation}{September 18th,
  2023}{Singapore}
\begin{document}

\title{Behind Recommender Systems: the Geography of the ACM RecSys Community}

\author{Lorenzo Porcaro}
\authornote{Both authors contributed equally to this research.}
\email{lorenzo.porcaro@ec.europa.eu}
\orcid{0000-0003-0218-5187}
\author{Jo\~{a}o Vinagre}
\authornotemark[1]
\email{joao.vinagre@ec.europa.eu}
\orcid{0000-0001-6219-3977}
\author{Pedro Frau}
\orcid{}
\author{Isabelle Hupont}
\orcid{0000-0002-9811-9397}
\author{Emilia G\'{o}mez}
\orcid{0000-0003-4983-3989}
\affiliation{%
  \institution{Joint Research Centre}
  \country{European Commission, Ispra, Italy, and Seville, Spain}
}

\renewcommand{\shortauthors}{Porcaro and Vinagre, et al.}

\maketitle

\section{Introduction}
The amount and dissemination rate of media content accessible online is nowadays overwhelming. 
Recommender Systems (RS) filter this information into manageable streams or feeds, adapted to our personal needs or preferences. 
It is of utter importance that algorithms employed to filter information do not distort or cut out important elements from our perspectives of the world. 
Under this principle, it is essential to involve diverse views and teams from the earliest stages of their design and development.
This has been highlighted, for instance, in recent European Union regulations such as the Digital Services Act~\cite{DSA}, via the requirement of risk monitoring, including the risk of discrimination, and the AI Act~\cite{AIact}, through the requirement to involve people with diverse backgrounds in the development of AI systems.

We look into the geographic diversity of the RS research community, specifically by analyzing the affiliation countries of the authors who contributed to the \textit{ACM Conference on Recommender Systems (RecSys)}\footnote{ACM RecSys webpage: \url{https://recsys.acm.org}} during the last 15 years. 
This study has been carried out in the framework of the \textit{Diversity in AI} -- DivinAI\footnote{DivinAI project website: \url{https://divinai.org}} project~\cite{Freire2021}, whose main objective is the long-term monitoring of diversity in AI forums through a set of indexes. 
While we have successfully applied these indexes to other fields~\cite{10041991}, this is the first in-depth analysis of the RecSys community.\footnote{All the data collected and analysed will be publicly available at \url{https://gitlab.com/humaint-ec_public/divinai-datasets}}

It is important to highlight that RecSys organisers have made several efforts to foster diversity in recent years. For instance, the figure of \textit{inclusion and accessibility chair} was introduced in RecSys 2018, and the initiatives to encourage women in RecSys started in 2014. 
Since 2021, in-person registration at discounted rates has been offered to participants from "economically developing" countries, as designated by ACM,\footnote{ACM's list of economically developing countries: \url{https://services.acm.org/public/qj/proflevel/countryListing.cfm}} based on World Bank's Gross National Income.
%

\section{Dataset}
We started the data collection process by querying Scopus,\footnote{Scopus bibliographic database: \url{https://www.scopus.com/}} looking for the RecSys proceedings from 2007 to 2022. 
For each paper we found in the proceedings, we extracted: 1) the list of authors; 2) the year of publication; and 3) each author's first affiliation.
%
After pre-processing the retrieved data to have more structured information (e.g. split 
authors' data into ordered lists), we focused on extracting for each author and conference year  
the country of 
affiliation, as provided in
the \textit{affiliation} field. 
%
%
%
In order to extract the country from affiliation, we used the \texttt{pycountry} library,\footnote{Python pycountry library: \url{https://pypi.org/project/pycountry/}} which gives the country names in the standard ISO-3166 format.
%
%
Afterwards, we double-checked the country information and corrected manually the data where needed, e.g., by 
adding the missing information in some cases where country 
could not be extracted from the metadata. From the total of 4277 entries so-collected (267 authors/conference year on average), manual correction was required in approximately 15\% of the entries.

\section{Metrics}
We replicate the method for computing the geographical diversity for each conference edition proposed in \cite{10041991}, and summarised hereafter.
Using the authors' affiliation countries, we compute the inverse Herfindahl-Hirschman Index (HHI), commonly used in economics to measure market concentration \cite{duch2022market}, and the Shannon Index (H') \cite{shannon1948mathematical}, commonly used in ecology to measure biodiversity. The higher the value of both metrics, the greater the diversity of countries represented at the conference.
HHI is computed following Eq. \eqref{metrics}--left, where $p_i=n_i/N$ is the proportion of elements from a given class $i$, i.e., the number of elements from this class $n_i$ divided by the total number of elements $N$, and $C$ is the number of different classes. 
The formula to compute the Shannon index is obtained 
in a similar fashion, but multiplying the proportion of elements by its logarithm (Eq. \eqref{metrics}--right). 
Note that here classes are the affiliation countries of contributors and that the total number of classes may vary over the years.

    \begin{equation}\label{metrics}
        HHI = {\sum\limits_{i=1}^{C}}p_i^2
        \qquad
        H'=-{\sum\limits_{i=1}^{C}}p_i\ln{p_i}
    \end{equation}

One of the advantages of using such indexes is that they are computed using both of the \textit{variety} of a set, i.e. its number of classes, and the \textit{balance}, i.e. the evenness of the distribution of elements across classes. 
As a result, we obtained an estimation of geographical diversity which is more compact and complete in comparison to the use of simple proportions, which, however, may be useful for a preliminary data exploration as shown in the next section.

\section{Results}
\begin{figure}[h!]\centering
\begin{subfigure}{0.45\textwidth}
\centering
\includegraphics[width = \textwidth]{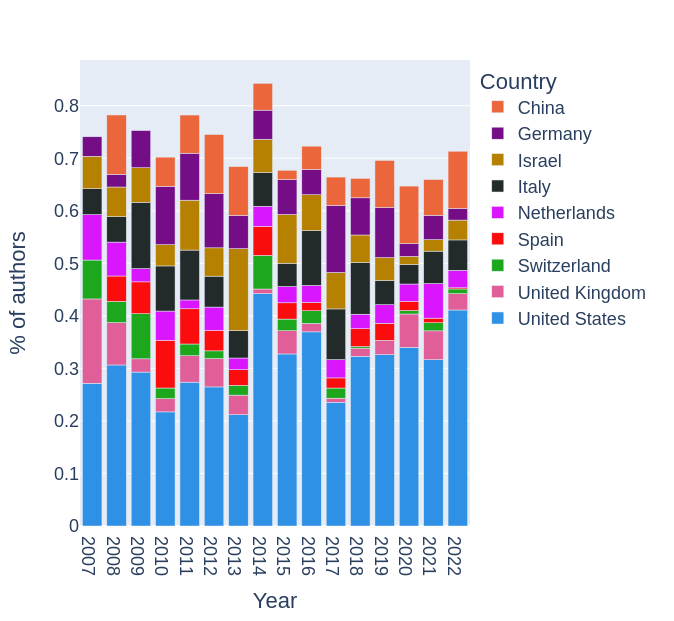}
\label{fig:left}
\end{subfigure}
\begin{subfigure}{0.45\textwidth}
\centering
\includegraphics[width = \textwidth]{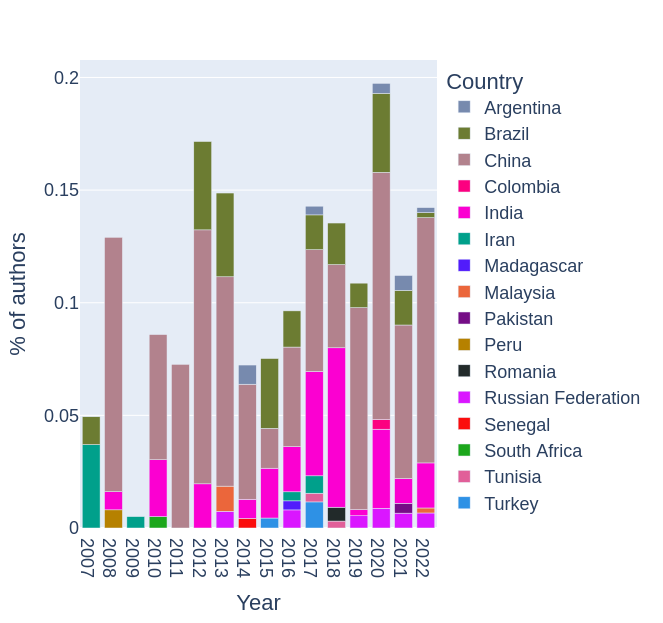}
\label{fig:right}
\end{subfigure}
\caption{Authors' presence distribution for top 5 countries (\textit{left}) and ``economically developing'' countries  (\textit{right}).}
\label{fig:1}
\end{figure}

We start our exploratory analysis by looking at the distribution of the authors' affiliation country in the several editions (Figure \ref{fig:1}).
In particular, we focus on two sets of countries.
The first set is formed by the countries that are at least in one edition of the conference among the top 5 in terms of authors' presence.
This set is formed by 9 countries: \textit{China}, \textit{Germany}, \textit{Israel}, \textit{Italy}, \textit{The Netherlands}, \textit{Spain}, \textit{Switzerland}, \textit{United Kingdom} and \textit{United States}. 
The second set is composed of the 139 countries for which ACM applies a special rate when registering for its conferences, part of the ones considered as ``economically developing''. 
%
%
%
The first set of 9 
countries have contributed to the conference with, in the lowest case, almost 65\% of the authors in 2020, with a peak of 85\% in 2014.
In this set, only China and Israel are not part of what are usually considered Western countries. 
In other words, 
for every RecSys conference in the last 15 years, at least 
2 out of 3 authors are affiliated with institutions from WEIRD (Western, Educated, Industrialised, Rich and Democratic) societies.
%
The USA is the country with the greatest 
presence, with a minimum of 21\% of authors in 2013, and a maximum of  44\% in 2014 when the conference was 
held in Silicon Valley. 

On the other side, we 
observe 
that the presence of authors from ``economically developing'' countries varied 
between less than 1\% in 2009 and around 20\% in 2020, the latter being the only edition in the history of the RecSys conference 
to be held entirely online. 
Within this set, most of the authors are based in the BRICS countries (Brazil, Russia, India, China, and South Africa), with China having a prominent role. 
Overall, there seems to be a trend where the percentage of authors from the top 5 countries set 
decreases slightly over the history of the conference, while the opposite is true for the "economically developing" countries, which show a slightly increasing trend.

A different perspective is given by the analysis of the diversity indexes presented in Figure \ref{fig:3}. 
By expanding to the list of all countries where RecSys authors have affiliations, the situation over the course of the years in terms of diversity seems to have been quite stable, perhaps with a very slight increasing trend in the Shannon index. 
The editions held in Minnesota (2007) and Silicon Valley (2014), emerge as the ones with the lowest geographical diversity, whilst Hong Kong (2013) and Como (2017) are the ones with 
the highest diversity.

\begin{table}[]
\begin{minipage}[b]{0.45\linewidth}
\centering

    \includegraphics[width = 0.9\linewidth]{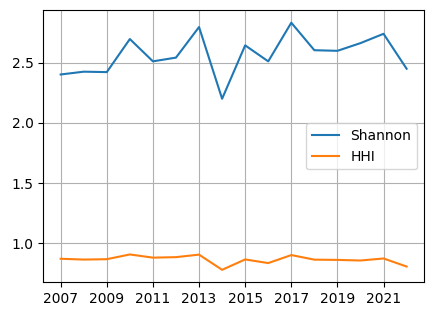}
    \captionof{figure}{HHI and Shannon diversity indexes for every RecSys edition.}
    \label{fig:3}
\end{minipage}\hfill
\begin{minipage}[b]{0.4\linewidth}
\centering
\resizebox{\linewidth}{!}{
\begin{tabular}{llrr}
\hline
\textbf{Year} & \textbf{Location}          & \textbf{Auth.} & \textbf{Countr.} \\ \hline
2007          & Minneapolis, USA           & 81                   & 17                          \\ \hline
2008          & Lausanne, Switzerland      & 124                  & 19                          \\ \hline
2009          & New York, USA              & 198                  & 20                          \\ \hline
2010          & Barcelona, Spain           & 198                  & 22                          \\ \hline
2011          & Chicago, USA               & 179                  & 20                          \\ \hline
2012          & Dublin, Ireland            & 204                  & 21                          \\ \hline
2013          & Hong Kong                  & 269                  & 28                          \\ \hline
2014          & Silicon Valey, USA         & 235                  & 23                          \\ \hline
2015          & Vienna, Austria            & 226                  & 26                          \\ \hline
2016          & Boston, USA                & 249                  & 30                          \\ \hline
2017          & Como, Italy                & 259                  & 34                          \\ \hline
2018          & Vancouver, Canada          & 325                  & 28                          \\ \hline
2019          & Copenhagen, Denmark        & 368                  & 30                          \\ \hline
2020          & Rio de Janeiro, Brazil     & 456                  & 33                          \\ \hline
2021          & Amsterdam, Netherlands     & 455                  & 33                          \\ \hline
2022          & Seattle, USA               & 451                  & 34                          \\ \hline
\end{tabular}}
\caption{RecSys location from 2007--2022, with number of participating authors and countries.}
\label{tab:conference_locations}
\end{minipage}
\end{table}

\section{Conclusions}
In light of the above observations, we may conclude that the geographical diversity of the RecSys community is still low, with more than two-thirds of authors coming from less than 10 countries. 
There seems to be a slight increase in the participation from economically less developed countries, however, it is premature to strongly relate this to recent efforts to make the conference more accessible. 
Other factors may come into play, such as the global increasing trend in AI research and the possibility of attending the conference online, at a discounted rate and without travel and accommodation expenses. 
Nevertheless, these measurements show that these efforts are all but unnecessary. We intend to provide more insights in the future, with a more detailed analysis accounting for the type of affiliation, topic diversity, gender and intersectional perspectives.

\begin{acks}
This work is partially supported by the HUMAINT programme (Human Behaviour and Machine Intelligence), Joint Research Centre, European Commission. The authors would like to thank the FAccTRec PC members and anonymous reviewers for their valuable feedback.
\end{acks}

\bibliographystyle{ACM-Reference-Format}
\bibliography{main}

\end{document}